\documentclass[12pt]{article}

\usepackage{amsmath}

\usepackage{amsfonts}

\usepackage{amssymb}

\usepackage{hyperref}

\usepackage[latin1]{inputenc}

\usepackage{graphics}

\usepackage{graphicx}

\usepackage{epsfig}

\usepackage{amsthm}

\usepackage{mathrsfs}

\usepackage{latexsym}

\usepackage{lineno}

\usepackage{latexsym}

\usepackage{float}

\usepackage{mathtools}

\usepackage[T1]{fontenc}

\usepackage{ae}

\usepackage{authblk}

\modulolinenumbers[5]

\newenvironment{destaque}{\begin{quotation}\small}{\end{quotation}}
\usepackage[figurename=Fig.]{caption}
\usepackage{subcaption}

\newcommand{\h}{\hspace{.5cm}}

\date{}

\begin{document}

\title{\bf Self-adjoint extensions for a $p^{4}$-corrected Hamiltonian of a particle on a finite interval}


\author[1]{B. B. Dilem\thanks{bernardob@ifes.edu.br}}
\author[2, 3]{J. C. Fabris\thanks{julio.fabris@cosmo-ufes.org}}
\author[3]{J. A. Nogueira\thanks{jose.nogueira@ufes.br}}
\affil[1]{Instituto Federal do Esp\'{\i}rito Santo -- Ifes\\ Alegre, Esp\'{\i}rito Santo, 29.520-000, Brasil}
\affil[2]{National Research Nuclear University MEPhI, Kashirskoe sh. 31, Moscow 115409, Russia}
\affil[3]{Universidade Federal do Esp\'{\i}rito Santo -- Ufes\\ Vit\'oria, Esp\'{\i}rito Santo, 29.075-910, Brasil}


\maketitle

\begin{abstract}
\begin{destaque}
In the present paper we deal with the issue of finding the self-adjoint extensions of a $p^4$-corrected Hamiltonian. The importance of this subject lies on the application of the concepts of quantum mechanics to the minimal-length scale scenario which describes an effective theory of quantum gravity. We work in a finite one dimensional interval and we give the explicit $U(4)$ parametrization that leads to the self-adjoint extensions. Once the parametrization is known, we can choose appropriate $U(4)$ matrices to model physical problems. As examples, we discuss the infinite square-well, periodic conditions, anti-periodic conditions and periodic conditions up to a prescribed phase. We hope that the parametrization we found will contribute to model other interesting physical situations in further works.     \\
\\
{\scriptsize PACS numbers: 04.60.-m, 02.30.Tb, 03.65.-w}\\
{\scriptsize Keywords: Minimal length; infinite well; self-adjoint extensions.}
\end{destaque}
\end{abstract}

\pagenumbering{arabic}


\section{Introduction}
\label{introduction}
\h In the last decades several papers dealing with applications of quantum mechanics to minimal-length scale scenarios have emerged, probably motivated by some results from theories using extra dimensions which suggest that the scale of the minimal length could be many order of magnitude greater than the Planck scale \cite{Arkani:1998, Randall:1999, Appelquist:2001, Hossenfelder1:2003}. In fact, the existence of a minimal length in nature is a general prediction of quantum gravity theories and its signatures must appear associated to transplanckian phenomena, hence outside the precision obtained in the present or near future experiments. However, extra dimensions can lower the Planck scale resulting in the possibility to detect signatures due to the existence of a minimal length.  Moreover, other results indicate that corrections on the spectra of some quantum mechanical system, due to the effects related to the minimal-length scenario, could be experimentally accessible at low energies. Although many of those paper have approached the problem by introducing, amongst some other things, a momentum fourth power correction in the Hamiltonian operator\footnote{In order to introduce a minimal-length scenario, which describes an effective theory of quantum gravity \cite{Hossenfelder:2006, Kober1:2012, Nouicer1:2012, Ong2:2018}, the Heisenberg's uncertainty principle is modified. There are many diferent proposals for modification, but almost all of them lead to a term of fourth power in the momentum operator as first correction to the kinetc energy of a non-relativistic particle, with the exception of the linear GUP (generalized uncertainty principle) \cite{Ali:2009} which leads to a term of third power.}, only very few of them had dealt with the question of its self-adjointness \cite{Vagenas:2015, Louko:2015, Farag:2016}. To cover this lacuna is the main purpose of this paper. It is important to remark that the self-adjointness of the Hamiltonian operator is a mathematical but essential issue in quantum mechanical physics, because it is this property that guarantees unitarity of the time-evolution operator and the existence of a dynamics, which assures, on the other hand, that the Cauchy problem
\begin{equation}
\frac{\partial \psi}{\partial t}=\hat{H} \psi ; \quad 
\psi|_{\, t=0}=\psi_{0},
\end{equation}
has a unique solution which conserves probability for all $\psi_{0}$ in the domain of $\hat{H}$ \cite{Gustafson-Sigal}. As it is well-known, unbounded self-adjoint operators can not be defined on all of Hilbert spaces. Consequently, we need to find suitable domains, that is, self-adjoint extensions of unbounded operators which are determined by boundary conditions. Hence, it is also important to remark that an operator is not completely defined only by its action, but also by its domain, which has to be specified by imposing appropriate (boundary) conditions on the functions on which it acts \cite{Bonneau:2001, Araujo:2004}. Therefore different boundary conditions lead to different operators (extensions) and thus to different physical situations.

The aim of this paper is to give an explicit and useful parametrization of the boundary conditions that specify the self-adjoint extensions of the $p^{4}$-corrected Hamiltonian. This Hamiltonian is described in next section. To reach this goal, we work in one dimension and with a free particle in the finite interval $x \in [-a/2, a/2]$. For this case, the parametrization is done by unitary matrices of dimension greater than those  corresponding to the half or full real line. Our choice for the one dimensional finite space was motived by a particular interest in the infinite square-well: In addition to its many applications, we had recently showed \cite{Oakes:2020} that the solutions obtained by taking the limit of a finite square-well are not the same found in the previous literature \cite{Nozari:2006, Pedram2:2010, Pedram:2016}. On that moment, it was not completely clear what was the additional conditions we should impose on the general solutions and, although our results at that point had gave us the correct clue, it is only by self-adjointness considerations, which is done in the third section, that we can properly define them.
 
Besides its obvious application to the infinite square-well, the finite interval can describe many other interesting situations by appropriate choice of the boundary conditions -- as we said: different conditions $\Rightarrow$ different physics. To illustrate this point, in the fourth section we explicitly present the matrices that lead to the self-adjoint boundary conditions for periodicity, anti-periodicity and periodicity up to a prescribed phase. We find their solutions and the corresponding energy spectrum. Although the eigenstates have very interesting properties under parity, time inversion and translations, they are not due to the minimal length correction and are the same exhibited by the ordinary quantum mechanics eigenstates, thus we left them for the Appendix \ref{App. 4}. Some of the explicit calculations were also left for the appendices.


\section{The $p^{4}$-corrected Hamiltonian and its self-adjoint extensions}
\label{MLS}

\h We consider the Hamiltonian operator
\begin{equation}
\label{corrected ham}
\hat{H} = -\frac{\hslash^{2}}{2m}\partial_{x}^{2}+\frac{\beta \hslash^{4}}{3m}\partial_{x}^{4}
\end{equation}
in the Hilbert space $\mathcal{H}= {\cal L}^{2}([-a/2,a/2],dx)$, where $a$ is a positive constant with dimension of length. If $\beta = 0$, then the Hamiltonian (\ref{corrected ham}) reduces to the ordinary quantum mechanical Hamiltonian of a free particle.

Just like in the ordinary case, the Halmitonian operator (\ref{corrected ham}) is an unbounded linear operator. In order to apply the von Neumann theorem \cite{Louko:2015, Bonneau:2001}, we compute its deficiency indices $(n_{+},n_{-})$, which can be defined, respectively, as the dimensions of the following subspaces:
\begin{eqnarray}
\label{def. spaces}
\mathcal{N_{+}}=\left\lbrace \psi \in \mathcal{D}(\hat{H}^{\dagger}), \quad \hat{H}^{\dagger}\psi=i\lambda_{+}\psi , \quad \lambda_{+}>0\right\rbrace ;\nonumber \\
\mathcal{N_{-}}=\left\lbrace \psi \in \mathcal{D}(\hat{H}^{\dagger}), \quad \hat{H}^{\dagger}\psi=i\lambda_{-}\psi , \quad \lambda_{-}<0\right\rbrace .
\end{eqnarray}
The calculations are shown in the Appendix \ref{App. 1} and the results are $n_{+}=n_{-}=4$, which means that this Hamiltonian operator has infinitely many self-adjoint extensions, parametrized by a unitary $4 \times 4$ matrix acting on the Hilbert space.

To find the boundary conditions that specify the self-adjoint extensions, we follow the same procedure of Ref. \cite{Louko:2015}, which basically consists in writing the symmetry condition $(\psi,\hat{H}\phi)=(\hat{H}\psi, \phi)$ as $C(u ,v)=0$, where $C(u ,v)$ is the sesquilinear form:
\begin{equation}
\label{sesq form}
C(u,v)=u^{\dagger}\cdot A\cdot v,
\end{equation}
with $A$ being a hermitian ($8 \times 8$) matrix that can be written in the form
\begin{eqnarray}
\label{A def.}
A=\left( \begin{array}{lr} G & O \\ O & -G \end{array} \right) ,
\end{eqnarray}
and find the maximal linear subspaces of $\mathbb{C}^{8}$ on which (\ref{sesq form}) vanishes. These subspaces give us the self-adjointness conditions for $\hat{H}$\footnote{Actually, if we have a linear subspace $V \subset \mathcal{H}$ on which
\begin{equation}
\label{sesquation}
u^{\dagger}\left( \begin{array}{lr} I & O \\ O & -I \end{array} \right)v = 0,
\end{equation}
for all $u, v \in V$, where $I$ is the identity matrix with appropriate dimension, then $u=v=\left( \begin{array}{c}
v_{1} \\ 
v_{2}
\end{array}  \right)$ implies $||v_{1}||=||v_{2}|| \Rightarrow v=\left( \begin{array}{c}
v_{1} \\ 
U.v_{1}
\end{array}  \right)$ for all $v \in V$, where $U$ is a constant unitary matrix with appropriate dimensions. So, the maximal linear subspaces $V_{U} \subset \mathcal{H}$ on which eq. (\ref{sesquation}) holds are
\begin{equation}
V_{U} = \left\lbrace v \in \mathcal{H} \left( \begin{array}{cc}
U & -I \\ 
0 & 0
\end{array} \right) v = 0 \right\rbrace .
\end{equation}
The matrix $A$ defined by (\ref{A def.}) can be written as \begin{equation}
A=(DP)^{\dagger}\left( \begin{array}{cc}
I & 0 \\ 
0 & -I
\end{array} \right)(DP),
\end{equation} where $D$ is a real positive diagonal matrix and $P$ is a unitary matrix. So, the maximal subspaces we are looking for are 
\begin{equation}
V_{U} = \left\lbrace v \in \mathcal{H} \left( \begin{array}{cc}
U & -I \\ 
0 & 0
\end{array} \right) (DP)v = 0 \right\rbrace .
\end{equation}}.

Integration by parts on the internal product $(\psi,\hat{H} \phi)$ shows that
\begin{eqnarray}
(\psi , \hat{H}\phi) &=& \int_{-a/2}^{a/2}\psi^{*}\left(-\frac{\hslash^{2}}{2m}\partial_{x}^{2}+\frac{\beta \hslash^{4}}{3m}\partial_{x}^{4}\right)\phi \: dx = \nonumber \\
&=& \left[-\frac{\hslash^{2}}{2m}\left(\psi^{*}\partial_{x}\phi -\partial_{x}\psi^{*}\phi \right)+ \frac{\beta\hslash^{4}}{3m} \left(\psi^{*}\partial_{x}^{3}\phi -\partial_{x}\psi^{*} \partial_{x}^{2}\phi +\partial_{x}^{2}\psi^{*} \partial_{x}\phi - \partial_{x}^{3}\psi^{*} \phi \right)\right]_{-a/2}^{a/2} + \nonumber \\ &&+ (\hat{H}\psi , \phi).
\end{eqnarray}
Hence, the symmetry condition can be written as
\begin{eqnarray}
\label{eq. fund}
\left[-\frac{\hslash^{2}}{2m}\left(\psi^{*}\partial_{x}\phi -\partial_{x}\psi^{*}\phi \right)+ \frac{\beta\hslash^{4}}{3m} \left(\psi^{*}\partial_{x}^{3}\phi -\partial_{x}\psi^{*} \partial_{x}^{2}\phi +\partial_{x}^{2}\psi^{*} \partial_{x}\phi - \partial_{x}^{3}\psi^{*} \phi \right)\right]_{-a/2}^{a/2} = \nonumber \\
= \frac{-i\beta \hslash^{4}}{3ma^{2}}\left[-i \, \frac{3a^{2}}{2\beta \hslash^{2}}\left(\psi^{*}\partial_{x}\phi -\partial_{x}\psi^{*}\phi \right)+ ia^{2} \left(\psi^{*}\partial_{x}^{3}\phi -\partial_{x}\psi^{*} \partial_{x}^{2}\phi +\partial_{x}^{2}\psi^{*} \partial_{x}\phi - \partial_{x}^{3}\psi^{*} \phi \right)\right]_{-a/2}^{a/2} =0 ,
\end{eqnarray}
which is the vanishing sesquilinear form we are looking for. Comparison with Eqs. (\ref{sesq form}) and (\ref{A def.}) gives us
\begin{eqnarray}
\label{G def.}
G=i\left(\begin{array}{ccrr} 0 & -\frac{3a^{2}}{2\beta \hslash^{2}} & 0 & 1 \\ \frac{3a^{2}}{2\beta \hslash^{2}} & 0 & -1 & 0 \\ 0 & 1 & 0 & 0 \\ -1 & 0 & 0 & 0\end{array} \right)
\end{eqnarray}
and the following $u$ and $v$:
\begin{eqnarray}
\label{u v def.}
u=\left(\begin{array}{lr}u_{+}\\u_{-}\end{array}\right), & v=\left(\begin{array}{lr}v_{+}\\v_{-}\end{array}\right), \end{eqnarray}
with
\begin{eqnarray}
\label{u+- v+- def.}
u_{\pm} = \left(\begin{array}{c} \psi(\pm a/2)\\ a\partial_{x}\psi(\pm a/2) \\ a^{2}\partial_{x}^{2}\psi(\pm a/2)
\\ a^{3}\partial_{x}^{3}\psi(\pm a/2)\end{array} \right) , & v_{\pm} = \left(\begin{array}{c} \phi(\pm a/2)\\ a\partial_{x}\phi(\pm a/2) \\ a^{2}\partial_{x}^{2}\phi(\pm a/2)
\\ a^{3}\partial_{x}^{3}\phi(\pm a/2)\end{array} \right).\end{eqnarray}
Further calculations (given in the Appendix \ref{App. 2}) lead us to the result
\begin{eqnarray}
\label{self-adj. cond.}
\left(\begin{array}{c}

\lambda_{+}\phi(+a/2) + i\lambda_{+}a\partial_{x}\phi(+a/2) - a^{2}\partial_{x}^{2}\phi(+a/2) - ia^{3}\partial_{x}^{3}\phi(+a/2) \\

\lambda_{+}\phi(-a/2)  - i\lambda_{+}a\partial_{x}\phi(-a/2) - a^{2}\partial_{x}^{2}\phi(-a/2) + ia^{3}\partial_{x}^{3}\phi(-a/2) \\

\lambda_{-}\phi(+a/2) - i\lambda_{-}a\partial_{x}\phi(+a/2) + a^{2}\partial_{x}^{2}\phi(+a/2) - ia^{3}\partial_{x}^{3}\phi(+a/2) \\

\lambda_{-}\phi(-a/2) + i\lambda_{-}a\partial_{x}\phi(-a/2) + a^{2}\partial_{x}^{2}\phi(-a/2) + ia^{3}\partial_{x}^{3}\phi(-a/2) \\

\end{array} \right)
= \nonumber \\
=U\left(\begin{array}{c}

\lambda_{+}\phi(+a/2) - i\lambda_{+}a\partial_{x}\phi(+a/2) - a^{2}\partial_{x}^{2}\phi(+a/2) + ia^{3}\partial_{x}^{3}\phi(+a/2)\\

\lambda_{+}\phi(-a/2) + i\lambda_{+}a\partial_{x}\phi(-a/2) - a^{2}\partial_{x}^{2}\phi(-a/2) - ia^{3}\partial_{x}^{3}\phi(-a/2) \\

\lambda_{-}\phi(+a/2) + i\lambda_{-}a\partial_{x}\phi(+a/2) + a^{2}\partial_{x}^{2}\phi(+a/2) + ia^{3}\partial_{x}^{3}\phi(+a/2) \\

\lambda_{-}\phi(-a/2)  -i\lambda_{-}a\partial_{x}\phi(-a/2) + a^{2}\partial_{x}^{2}\phi(-a/2) - ia^{3}\partial_{x}^{3}\phi(-a/2) \\

\end{array} \right),
\end{eqnarray}
where
\begin{equation}
\label{lamb def}
\lambda_{\pm}=\frac{\sqrt{1+\left(\frac{4\beta \hslash^{2}}{3a^{2}}\right)^{2}}\pm 1}{4\beta\hslash^{2}/3a^{2}},
\end{equation}
and $U \in U(4)$ is the unitary $4\times 4$ matrix that specifies self-adjoint extension.

Two important remarks are in order now: First, all the self-adjoint extensions of $\hat{H}$, for the one dimensional finite interval, are given by (\ref{self-adj. cond.}). Thus, if the question is to find out if some known boundary conditions are enough to ensure self-adjointness, or if some known set of eigenfunctions belongs to a self-adjoint $\hat{H}$ domain, the answer can be obtained by finding the $U(4)$ matrix that leads to that boundary conditions or set of eigenfunctions -- or prove there is no such matrix. In the next section we will find the matrices that lead to some known boundary conditions and set of eigenfunctions, related to the infinite square-well, proving they are associated with self-adjoint Hamiltonian extensions.

Second, all $U(4)$ possible choice for (\ref{self-adj. cond.}) leads to a self-adjoint Hamiltonian. Thus, there are infinitely many models to work with and all we have to do is to pick a $U(4)$ matrix. Some groups of matrix could be classified in subfamilies according to their proprieties or what kind of conditions they lead to. In the fourth section, we will explore some different models by choosing different matrices.

Multiplying the Eq. (\ref{self-adj. cond.}) by $\lambda_{-}$ and noting that $\lambda_{-}\lambda_{+} = 1$, we have, up to $\cal{O}(\beta$), 
\begin{eqnarray}
\label{beta_zero}
\left(\begin{array}{c}

\phi(+a/2) + ia\partial_{x}\phi(+a/2) - \frac{2}{3}\hbar^{2}\beta \partial_{x}^{2} \phi(+a/2) - i \frac{2}{3}\hbar^{2} a \beta \partial_{x}^{3} \phi(+a/2) \\

\phi(-a/2)  - ia\partial_{x}\phi(-a/2) -\frac{2}{3}\hbar^{2}\beta \partial_{x}^{2} \phi(-a/2) + i \frac{2}{3}\hbar^{2} a \beta \partial_{x}^{3} \phi(-a/2)\\

\frac{2}{3}\hbar^{2}\beta \partial_{x}^{2} \phi(+a/2) - i \frac{2}{3}\hbar^{2} a \beta \partial_{x}^{3} \phi(+a/2) \\

\frac{2}{3}\hbar^{2}\beta \partial_{x}^{2} \phi(-a/2) + i \frac{2}{3}\hbar^{2} a \beta \partial_{x}^{3} \phi(-a/2) \\

\end{array} \right)
= \nonumber \\
=U\left(\begin{array}{c}

\phi(+a/2) - ia\partial_{x}\phi(+a/2) -\frac{2}{3}\hbar^{2}\beta \partial_{x}^{2} \phi(+a/2) + i \frac{2}{3}\hbar^{2} a \beta \partial_{x}^{3} \phi(+a/2) \\

\phi(-a/2) + ia\partial_{x}\phi(-a/2) -\frac{2}{3}\hbar^{2}\beta \partial_{x}^{2} \phi(-a/2) - i \frac{2}{3}\hbar^{2} a \beta \partial_{x}^{3} \phi(-a/2) \\

\frac{2}{3}\hbar^{2}\beta \partial_{x}^{2} \phi(+a/2) + i \frac{2}{3}\hbar^{2} a \beta \partial_{x}^{3} \phi(+a/2) \\

\frac{2}{3}\hbar^{2}\beta \partial_{x}^{2} \phi(-a/2) - i \frac{2}{3}\hbar^{2} a \beta \partial_{x}^{3} \phi(-a/2) \\

\end{array} \right),
\end{eqnarray}
which can be rewritten as
\begin{eqnarray}
\label{beta_zero1}
\left(\begin{array}{c}

\varphi_{1}^{+} \\

\varphi_{2}^{+}  \\

\end{array} \right) 
=\left( \begin{array}{rr}
A & B \\ 
C & D \\ 
\end{array} \right) 
\left(\begin{array}{c}

\varphi_{1}^{-} \\

\varphi_{2}^{-} \\

\end{array} \right),
\end{eqnarray}
where
\begin{eqnarray}
\label{varphi1}
\varphi_{1}^{\pm} :=
\left(\begin{array}{c}
\phi(+a/2) \pm ia\partial_{x}\phi(+a/2) - \frac{2}{3}\hbar^{2}\beta \partial_{x}^{2} \phi(+a/2) \mp i \frac{2}{3}\hbar^{2} a \beta \partial_{x}^{3} \phi(+a/2) \\
\phi(-a/2)  \mp ia\partial_{x}\phi(-a/2) - \frac{2}{3}\hbar^{2}\beta \partial_{x}^{2} \phi(-a/2) \pm i \frac{2}{3}\hbar^{2} a \beta \partial_{x}^{3} \phi(-a/2) \\
\end{array} \right),
\end{eqnarray}

\begin{eqnarray}
\label{varphi2}
\varphi_{2}^{\pm} :=
\left(\begin{array}{c}
\frac{2}{3}\hbar^{2}\beta \partial_{x}^{2} \phi(+a/2) \mp i \frac{2}{3}\hbar^{2} a \beta \partial_{x}^{3} \phi(+a/2) \\
\frac{2}{3}\hbar^{2}\beta \partial_{x}^{2} \phi(-a/2) \pm i \frac{2}{3}\hbar^{2} a \beta \partial_{x}^{3} \phi(-a/2) \\
\end{array} \right)
\end{eqnarray}
and
\begin{equation}
U=\left( \begin{array}{cc}
A & B \\ 
C & D
\end{array}  \right).
\end{equation}
Now, taking the limit $\beta \rightarrow 0$ in Eq. (\ref{beta_zero1}) we have $C \varphi_{0}^{-} = 0$, where
\begin{eqnarray}
\label{varphi1}
\varphi_{0}^{\pm} :=
\left(\begin{array}{c}
\phi(+a/2) \pm ia\partial_{x}\phi(+a/2) \\
\phi(-a/2)  \mp ia\partial_{x}\phi(-a/2) \\
\end{array} \right)
\end{eqnarray}
represents the non vanishing terms of Eq. (\ref{beta_zero}) when $\beta = 0$ (which are the same terms that we have in the ordinary case). Therefore, if $C^{-1}$ exists, that is, $C$ is an invertible matrix, then $\varphi_{0}^{-} = 0$. Consequently, $\phi(\pm a/2) = 0$ and $\partial_{x} \phi(\pm a/2) = 0$. However, those boundary conditions imposed to the ordinary case ($\beta = 0$) lead to trivial solution $\phi(x) = 0$. If we rewrite Eq. (\ref{beta_zero1}) as
\begin{eqnarray}
\label{beta_zero2}
\left( \begin{array}{rr}
A^{\dagger} & C^{\dagger} \\ 
B^{\dagger} & D^{\dagger} \\ 
\end{array} \right) 
\left(\begin{array}{c}

\varphi_{1}^{+} \\

\varphi_{2}^{+}  \\

\end{array} \right) 
=
\left(\begin{array}{c}

\varphi_{1}^{-} \\

\varphi_{2}^{-} \\

\end{array} \right),
\end{eqnarray}
we can follow the same development to conclude that, since Eq. (\ref{beta_zero2}) leads to $B^{\dagger}\varphi_{0}^{+}=0$ in the limit $\beta \rightarrow 0$, if $B^{-1}$ exists, then $\phi(\pm a/2) = 0$ and $\partial_{x} \phi(\pm a/2) = 0$ in this same limit, what leads again to the trivial solution $\phi(x)=0$.\\
Otherwise, if $B$ and $C$ are both null matrices\footnote{In fact, due to the unitarity of $U$, $B=0$ $\Leftrightarrow$ $C=0$.}, then $A$ will be a unitary $2\times 2$ matrix, as a consequence of the unitarity of $U$. Then, in the limit $\beta \rightarrow 0$ we have:
\begin{equation}
\label{beta_zero3}
\varphi_{0}^{+}=A.\varphi_{0}^{-},
\end{equation}
which is the parametrization of the ordinary case\footnote{Actually, Eq. (\ref{beta_zero3}) is also obtained if $B\neq 0$ and $C\neq 0$, but not necessarily with unitary $A$}:
\begin{eqnarray}
\left(\begin{array}{c}
\phi(+a/2) + ia\partial_{x}\phi(+a/2) \\
\phi(-a/2)  - ia\partial_{x}\phi(-a/2) \\
\end{array} \right)
=U(2).\left(\begin{array}{c}
\phi(+a/2) - ia\partial_{x}\phi(+a/2) \\
\phi(-a/2)  + ia\partial_{x}\phi(-a/2) \\
\end{array} \right).
\end{eqnarray}
Note that when $\phi(x)$ can be written as a perturbative series in power of $\beta$, $\phi(x) = \sum_{N=0}^{\infty} \beta^{N} \phi_{N}(x)$, the boundary conditions obtained in the limit ${\beta \to 0}$ are satisfied by the solution of the ordinary case, $\lim_{\beta \to 0} \left[ \partial_{x}^{n} \phi(\pm a/2) \right] = \partial_{x}^{n} \phi_{0}(\pm a/2)$. Therefore, if the boundary conditions in $p^{4}$-corrected theory are not the same as uncorrected theory the wave function of the $p^{4}$-corrected theory can not be written as a perturbative power series in $\beta$.

\section{The infinite square-well}
\label{ISW}

\h  In ordinary quantum mechanics, the boundary conditions of the infinite square-well state that the wave function of the particle vanishes at the both edges of the well:
\begin{equation}
\psi(-a/2)=\psi(a/2)=0 ,
\end{equation} 
what leads to the well-parity-defined eigenfunctions
\begin{equation}
\label{ord. sol.}
\left\lbrace 
\begin{array}{l}
\psi(x)=\sqrt{\frac{2}{a}} \cos(\frac{n\pi x}{a}),\quad $n odd$ \\
\psi(x)=\sqrt{\frac{2}{a}} \sin(\frac{n\pi x}{a}),\quad $n even$
\end{array} \right. .
\end{equation}
In a minimal-length scenario, the fourth order derivative implies that we have additional boundary conditions to impose on the wave function and its derivatives. Despite of that, and since (\ref{ord. sol.}) mathematically satisfies the time-independent Schroedinger equation
\begin{equation}
\label{tise}
\hat{H} \psi = E \psi,
\end{equation}
\begin{equation}
-\frac{\hslash^{2}}{2m}\partial_{x}^{2}\psi+\frac{\beta \hslash^{4}}{3m}\partial_{x}^{4}\psi = E\psi,
\end{equation}
with $E = E_{n} = \frac{n^{2} \pi^{2} \hbar^{2}}{2ma^{2}} + \beta \frac{n^{4} \pi^{4} \hbar^{4}}{3ma^{4}}$, some authors have bypassed the problem of finding the appropriate boundary conditions and have considered (\ref{ord. sol.}) as the correct eigenfunctions for the minimal-length  infinite square-well \cite{Nozari:2006, Pedram2:2010, Pedram:2016,Esguerra:2016}. In fact, if we choose the $U(4)$ matrix
\begin{equation}
U_{4} = \left( \begin{array}{rrrr}
-1 & 0 & 0 & 0 \\ 
0 & -1 & 0 & 0 \\ 
0 & 0 & -1 & 0 \\ 
0 & 0 & 0 & -1
\end{array} \right) 
\end{equation}
in the parametrization (\ref{self-adj. cond.}), we will get the following conditions:
\begin{equation}
\psi(-a/2)=\psi(a/2)=\partial_{x}^{2}\psi(-a/2)=\partial_{x}^{2}\psi(a/2)=0,
\end{equation}
which really leads to the eigenfunctions (\ref{ord. sol.}). Hence, (\ref{ord. sol.}) do came from conditions that makes the $p^{4}$ corrected Hamiltonian self-adjoint, but are they the true boundary conditions for the infinite square-well? Apparently, the answer is no.

In a recent paper we consider the infinite square-well as a limiting case of the finite square-well and found a different set of eigenfunction - with continuous first derivative at the walls - as solution \cite{Oakes:2020}. At that point, we did not demonstrate the self-adjointness of the Hamiltonian operator, but now its a straightforward calculation to verify that the $U(4)$ matrix
\begin{equation}
\label{U paramet}
U=\left( 
\begin{array}{rrrr}
0 & 0 & -1 & 0\\
0 & 0 & 0 & -1\\
-1 & 0 & 0 & 0\\
0 & -1 & 0 & 0
\end{array}
\right)
\end{equation}
applied to (\ref{self-adj. cond.}) gives the following conditions:
\begin{equation}
\label{cond. cont}
\left\lbrace
\begin{array}{c}
\lambda_{+}\phi (\pm a/2)=-\lambda_{-}\phi (\pm a/2) \\
\lambda_{+}\partial_{x}\phi (\pm a/2)=-\lambda_{-}\partial_{x}\phi (\pm a/2)
\end{array}
\right. \qquad \Rightarrow \qquad
\left\lbrace
\begin{array}{c}
\phi (a/2)=\phi (-a/2)=0 \\
\partial_{x}\phi (a/2)=\partial_{x}\phi (-a/2)=0
\end{array}
\right. ,
\end{equation}
which leads to the same results we found, that is, the following eigenfunctions of opposite parity\footnote{The constants $A_{k}$ and $B_{k}$ have been defined by
\begin{eqnarray}
A_{k}&=&\cos(ka/2)\cosh(k'a/2)\left[ \cosh^{2}(k'a/2) \left(\frac{a}{2}+\frac{(k'^{2}-3k^{2})\sin(ka/2)\cos(ka/2)}{k(k^{2}+k'^{2})}\right) \right. + \nonumber\\
&& + \left. \cos^{2}(ka/2) \left(\frac{a}{2}+\frac{(k^{2}-3k'^{2})\sinh(k'a/2)\cosh(k'a/2)}{k'(k'^{2}+k^{2})}\right) \right]^{-1/2}
\end{eqnarray}
and
\begin{eqnarray}
B_{k}&=&\sin(ka/2)\sinh(k'a/2)\left[ \sinh^{2}(k'a/2) \left(\frac{a}{2}-\frac{(k'^{2}-3k^{2})\sin(ka/2)\cos(ka/2)}{k(k^{2}+k'^{2})}\right) \right. + \nonumber\\
&& - \left. \sin^{2}(ka/2) \left(\frac{a}{2}-\frac{(k^{2}-3k'^{2})\sinh(k'a/2)\cosh(k'a/2)}{k'(k'^{2}+k^{2})}\right) \right]^{-1/2},
\end{eqnarray}
in order to normalize the wave functions.}:
\begin{equation}
\label{Asol}
\psi_{A_{k}} = A_{k}\left[\frac{\cos(kx)}{\cos(ka/2)}-\frac{\cosh(k'x)}{\cosh(k'a/2)}\right]
\end{equation}
and
\begin{equation}
\label{Bsol}
\psi_{B_{k}} = B_{k}\left[\frac{\sin(kx)}{\sin(ka/2)}-\frac{\sinh(k'x)}{\sinh(k'a/2)}\right],
\end{equation}
and the following conditions that define the energy spectrum:
\begin{equation}
\label{even spectrum}
k\tan(ka/2)+k'\tanh(k'a/2)=0,
\end{equation}
to the even eigenfunctions, and
\begin{equation}
\label{odd spectrum}
k\cot(ka/2)-k'\coth(k'a/2)=0
\end{equation}
to the odd ones, with
\begin{equation}
k=\frac{\sqrt{\sqrt{1+\frac{16}{3}\beta mE}-1}}{2\hslash \sqrt{\beta /3}}, \qquad k'=\frac{\sqrt{\sqrt{1+\frac{16}{3}\beta mE}+1}}{2\hslash \sqrt{\beta /3}}.
\end{equation}
\\

It is important to stress that in both cases the ordinary infinite square-well is recovered in the limit ${\beta \rightarrow 0}$ since:
\begin{itemize}
\item the energy equations (\ref{even spectrum}) and (\ref{odd spectrum}) become the energy equations of the ordinary infinite square-well,\

\item $ \left[ \lim_{\beta \to 0} \left\{ \psi_{A_{k}, B_{k}}(x) \right\} \right]_{x = \pm a/2} = \psi^{0}_{A_{k_{0}}, B_{k_{0}}}(\pm a/2) = 0$,

\item $ \left[ \lim_{\beta \to 0}  \left\{ \partial_{x}\psi_{A_{k}, B_{k}}(x) \right\} \right]_{x = \pm a/2} = \partial_{x}\psi^{0}_{A_{k_{0}}, B_{k_{0}}}(\pm a/2) \neq 0$,
\end{itemize}
where $\psi^{0}_{A_{k_{0}}, B_{k_{0}}(x)}$ are the solutions of the ordinary infinite square-well.
In the second case, the matrix 
$\left(
\begin{array}{rr}
-1 & 0\\
 0 & -1 
\end{array}
\right) $ is invertible and thus the boundary conditions $\phi(\pm a/2) = 0$ and $\partial_{x}\phi(\pm a/2) = 0$ are held in the limit ${\beta \rightarrow 0}$, which are not the boundary conditions of the ordinary infinite square-well. So, it is expected that the solutions of the infinite square-well in minimal-length scenario in the second case are not given by a perturbative series in powers of $\beta$. A quick glance at Eqs. (\ref{Asol}) and (\ref{Bsol}) shows that this is the case.




\section{Some other models from different choices of the Hamiltonian self-adjoint extension}
\label{SOMFDCOTHSAE}

\h In this section we explore some other possible choices to (\ref{self-adj. cond.}): periodic conditions, anti-periodic conditions, and periodic conditions up to a phase factor. We expose their solutions and energy spectrum. Their symmetry properties under parity, time inversion and translation are very interesting, but they are similar to the ordinary quantum mechanical case, so we left them to the Appendix \ref{App. 4}. 

\subsection{Periodic conditions}

Starting with the Eq. (\ref{self-adj. cond.}), a possible choice is
\begin{equation}
\label{U paramet2}
U=\left( 
\begin{array}{rrrr}
0 & 1 & 0 & 0\\
1 & 0 & 0 & 0\\
0 & 0 & 0 & 1\\
0 & 0 & 1 & 0
\end{array}
\right), 
\end{equation}
which takes us to the following boundary conditions:
\begin{equation}
\label{ring cond.}
\left\lbrace \begin{array}{c}
\phi(+a/2)=\phi(-a/2)\\
\partial_{x}\phi(+a/2)=\partial_{x}\phi(-a/2)\\
\partial_{x}^{2}\phi(+a/2)=\partial_{x}^{2}\phi(-a/2)\\
\partial_{x}^{3}\phi(+a/2)=\partial_{x}^{3}\phi(-a/2)
\end{array} \right. .
\end{equation}
The full periodic conditions above suggests that we might identify the both edges of the interval $[-a/2, a/2]$ as the same physical point and, doing so, we may interpret the solutions of this mathematical model as the wave function of a particle trapped on a ring of length $a$. Returning to the eigenfunctions of the Hamiltonian operator (\ref{tise}):
\begin{equation}
\phi(x) = Ae^{ikx}+Be^{-ikx}+Ce^{k'x}+De^{-k'x}, \nonumber
\end{equation}
we are lead,  from Eq. (\ref{ring cond.}), to the following solutions (all of them with $E\geq0$):
\begin{equation}
\label{ring sol.}
\phi_{n}^{\pm}(x)=\frac{e^{\pm i \left( 2n\pi x/a-\theta \right)}}{\sqrt{a}} ; \qquad E_{n}^{\pm}=\frac{\hslash^{2}}{2m}\left(\frac{2n\pi}{a}\right)^{2}+\frac{\beta \hslash^{4}}{3m}\left(\frac{2n\pi}{a}\right)^{4},
\end{equation}
where $n \in \mathbb{N}^{*}$ (in order to satisfy the boundary conditions) and $\theta \in [0,2\pi)$ is a phase factor we introduce in order to stress that, in a ring, the origin does not have any special meaning over any other point -- note that the introduction of $\theta$ have the effect of translating the wave function for a distance $0\leq \delta=a\theta /(2n\pi)\leq a/n$, which is enough to make the maximum displacement to match the distance from which it starts to repeat itself. By the above solution, we can see that each eigenvalue is doubly degenerate, since, for fixed $n$, $E_{n}^{+}=E_{n}^{-}=E_{n}$. The only exception is the ground state, given -- except for a constant phase -- by
\begin{equation}
\label{ring sol. II}
\phi_{0}(x)=\frac{1}{\sqrt{a}},
\end{equation} 
which have null energy.

\subsection{Anti-periodic conditions}

If, instead of (\ref{U paramet2}), we choose
\begin{equation}
\label{U paramet3}
U=\left( 
\begin{array}{rrrr}
0 & -1 & 0 & 0\\
-1 & 0 & 0 & 0\\
0 & 0 & 0 & -1\\
0 & 0 & -1 & 0
\end{array}
\right),
\end{equation}
in (\ref{self-adj. cond.}), we are taken to the following boundary conditions:
\begin{equation}
\label{anti-ring cond.}
\left\lbrace \begin{array}{c}
\phi(+a/2)=-\phi(-a/2)\\
\partial_{x}\phi(+a/2)=-\partial_{x}\phi(-a/2)\\
\partial_{x}^{2}\phi(+a/2)=-\partial_{x}^{2}\phi(-a/2)\\
\partial_{x}^{3}\phi(+a/2)=-\partial_{x}^{3}\phi(-a/2)
\end{array} \right. ,
\end{equation}
which leads to the solutions
\begin{equation}
\label{anti-ring sol.}
\phi_{n}^{\pm}(x)=\frac{e^{ i \left[\left(1 \pm 2n \right) \pi x/a\mp \theta \right]}}{\sqrt{a}} ; \qquad E_{n}^{\pm}=\frac{\hslash^{2}}{2m}\left(\frac{2n\pi \pm \pi}{a}\right)^{2}+\frac{\beta \hslash^{4}}{3m}\left(\frac{2n\pi \pm \pi}{a}\right)^{4},
\end{equation}
with $n \in \mathbb{N}$ (as a consequence of (\ref{anti-ring cond.})) and $\theta \in [0,2\pi)$. Although similar to the periodic case, now we see that not only the excited states are degenerated, but also the ground state (represented by $\phi_{0}(x)$ and by $\phi_{1}^{-}(x)$). Indeed, we see that $E_{n}^{+}=E_{n+1}^{-}$ and, therefore, the states $\phi_{n}^{+}$ and $\phi_{n+1}^{-}$, although different, have the same energy.

\subsection{Periodic conditions up to a phase factor}

The periodic conditions up to a phase factor generalizes the both cases above, connecting them by a single continuous parameter $\varphi \in [0,2\pi)$. For this purpose, we choose for (\ref{self-adj. cond.}) the parametrization
\begin{equation}
\label{U paramet4}
U=e^{i\varphi} \, \left( 
\begin{array}{rrrr}
0 & 1 & 0 & 0\\
1 & 0 & 0 & 0\\
0 & 0 & 0 & 1\\
0 & 0 & 1 & 0
\end{array}
\right).
\end{equation}
From (\ref{U paramet4}), we obtain
\begin{equation}
\label{phase ring cond.}
\left\lbrace \begin{array}{c}
\phi(+a/2)=e^{i\varphi}.\phi(-a/2)\\
\partial_{x}\phi(+a/2)=e^{i\varphi}.\partial_{x}\phi(-a/2)\\
\partial_{x}^{2}\phi(+a/2)=e^{i\varphi}.\partial_{x}^{2}\phi(-a/2)\\
\partial_{x}^{3}\phi(+a/2)=e^{i\varphi}.\partial_{x}^{3}\phi(-a/2)
\end{array} \right. .
\end{equation}
These conditions lead us to the following solutions:
\begin{equation}
\label{phase ring sol.}
\phi_{n}^{\pm}(x)=\frac{e^{ i \left[\left(\varphi \pm 2n\pi \right) x/a\pm \theta \right]}}{\sqrt{a}} ; \qquad E_{n}^{\pm}=\frac{\hslash^{2}}{2m}\left(\frac{2n\pi \pm \varphi}{a}\right)^{2}+\frac{\beta \hslash^{4}}{3m}\left(\frac{2n\pi \pm \varphi}{a}\right)^{4},
\end{equation}
that are non-degenerate -- except if $\varphi = 0$ or $\varphi = \pi$, which make us to return to the periodic and anti-periodic conditions, respectively. 



\section{Conclusions and further perspectives}
\label{Concl}

\h We have studied a free non-relativistic quantum particle, on a finite interval, described by a Hamiltonian with a $p^{4}$ correction. The correction is part of the description of quantum mechanics in a minimal-length scale scenario, and the importance of our work lies in its applications to this field. At first, we had focus our attention on the question of Hamiltonian self-adjointness. Our main result on this subject was to describe explicitly the $U(4)$ family of boundary conditions that parametrizes the self-adjoint extensions of the Hamiltonian operator. The description we found brings the minimal-length parameter $\beta$ inside itself, through the constant $\lambda_{\pm}$ (Eq. (\ref{lamb def})), but, in the situations we had worked with, its presence was excluded due to the boundary conditions. Thus, it would be important to investigate, in a further work, what kind of systems have $\lambda_{\pm}$, and hence $\beta$, on its boundary conditions.

In the third section, we had turned our attention to the infinite square-well problem. Due to the $p^4$ correction, the continuity of the wave functions at the edge of the well is not enough to ensure the self-adjointess of the Hamiltonian operator. The eigenfunctions and energy equation obtained by taking the infinite square-well as a limit case of the finite square-well suggest that the continuity of the first derivative at the both edges is the missing boundary condition to be imposed on the solutions. By finding the appropriate $U(4)$ matrix, we have finally proved that it was really the case: The vanishing of both the wave function and its first derivative at the walls of the square well guaranties a self-adjoint Hamiltonian and also leads to the same results that we obtained by the limiting process. We verified that there are self-adjoint conditions that lead to the same eigenfunctions found in the previous literature \cite{Nozari:2006, Pedram2:2010, Pedram:2016,Esguerra:2016}, which are the same of ordinary quantum mechanics. However, those conditions include the vanishing of the second derivative at the infinite square-well walls. Further investigations would be necessary to find if both of those solutions have some physical meaning and which of them describes an infinite square-well in a minimal-length scenario.

Finally, it is important to remark that the knowledge of the self-adjoint parametrization we found in the second section enables us to study, on a solid mathematical basis, many different physical systems in this same context, selected by a $U(4)$ matrix choice that leads to the appropriate boundary conditions. As examples, we found the eigenfunctions and corrected energy spectrum for periodic, anti-periodic and periodic up to a phase conditions. We expect that our results will be applied to other physical models as a natural continuation of the present work.


\section*{Acknowledgements}

\h Bernard Brunoro Dilem would like to thank Daniel Franco (Universidade Federal de Vi\c{c}osa) for their conversations about self-adjoint extensions of operators during his post-doctoral stage.

The authors would like to thank Giuseppe Dito (Universit\'e de Bourgogne) for useful discussions and suggestions, and FAPES, CAPES and CNPq (Brazil) for financial support.
\\


\appendix
\section{Appendix: Calculating the deficiency indices}
\label{App. 1}

\h This appendix shows the calculations that lead to the deficiency indices of Section \ref{ISW}. \\
\\
Starting with the equation
\begin{equation}
\label{def. eq.}
\hat{H}^{\dagger}\phi_{\pm}=\pm i\lambda \phi_{\pm} 
\end{equation}
and assuming that $\hat{H}$ is formally self-adjoint, that is, assuming that $\hat{H}$ and $\hat{H}^{\dagger}$ have the same form, the next step is to solve the equation
\begin{equation}
\left(-\frac{\hslash^{2}}{2m}\partial_{x}^{2}+\frac{\beta \hslash^{4}}{3m}\partial_{x}^{4} \right)\phi_{\pm}=\pm i\lambda \phi_{\pm}.
\end{equation}
If we write
\begin{equation}
\phi_{\pm}=e^{k_{\pm}x},
\end{equation}
then we will have:
\begin{equation}
-\frac{\hslash^{2}}{2m}k_{\pm}^{2}+\frac{\beta \hslash^{4}}{3m}k_{\pm}^{4} =\pm i\lambda \quad \Leftrightarrow \quad
\frac{2\beta \hslash^{2}}{3}k_{\pm}^{4}- k_{\pm}^{2}  \mp  \frac{2mi\lambda}{\hslash^{2}} =0 ,
\end{equation}
whose solutions are:
\begin{equation}
k_{\pm}^{2}=\frac{1+\sqrt{1\pm 16im\beta \lambda/3}}{4\beta \hslash^{2}/3}
\end{equation}
and
\begin{equation}
k_{\pm}^{'2}=\frac{1-\sqrt{1\pm 16im\beta \lambda/3}}{4\beta \hslash^{2}/3}.
\end{equation}
The term
\begin{equation}
\sqrt{1\pm 16im\beta \lambda/3}
\end{equation}
is a complex number and, then, can be written as 
\begin{equation}
z=a+ib.
\end{equation}
Now, we have
\begin{equation}
\label{zz}
|z|^{2}=a^{2}+b^{2}=z^{*}z=\sqrt{1^{2}+(16m\beta \lambda/3)^{2}}
\end{equation}
and
\begin{equation}
\label{z^2}
z^{2}=a^{2}-b^{2}+2iab=1\pm 16im\beta \lambda/3 \quad \Rightarrow \quad \left\lbrace
\begin{array}{l}
a^{2}-b^{2}=1 \\
2ab=\pm 16m\beta \lambda/3
\end{array} \right. .
\end{equation}
Adding and subtracting the real part of (\ref{z^2}) from (\ref{zz}), we find
\begin{equation}
|a|=\sqrt{\frac{\sqrt{1^{2}+(16m\beta \lambda /3)^{2}}+1}{2}}
\end{equation}
and
\begin{equation}
|b|=\sqrt{\frac{\sqrt{1^{2}+(16m\beta \lambda /3)^{2}}-1}{2}},
\end{equation}
which implies that
\begin{equation}
\label{+}
\sqrt{1+ 16im\beta \lambda/3}=\pm \left( \sqrt{\frac{\sqrt{1^{2}+(16m\beta \lambda /3)^{2}}+1}{2}} + i \sqrt{\frac{\sqrt{1^{2}+(16m\beta \lambda /3)^{2}}-1}{2}} \right)
\end{equation}
and
\begin{equation}
\label{-}
\sqrt{1- 16im\beta \lambda/3}=\pm \left( \sqrt{\frac{\sqrt{1^{2}+(16m\beta \lambda /3)^{2}}+1}{2}} - i \sqrt{\frac{\sqrt{1^{2}+(16m\beta \lambda /3)^{2}}-1}{2}} \right),
\end{equation}
where the signal of the terms in parentheses were defined in agreement with the imaginary part of (\ref{z^2}).\\
\\
Now, we can write:
\begin{equation}
\phi_{\pm}=A_{\pm}e^{+ k_{\pm}x}+B_{\pm}e^{- k_{\pm}x}+C_{\pm}e^{+ k'_{\pm}x}+D_{\pm}e^{- k'_{\pm}x},
\end{equation}
with
\begin{equation}
k_{\pm}=\sqrt{\frac{1+(|a|\pm i|b|)}{4\beta \hslash^{2}/3}}=\sqrt{\frac{(1+|a|)\pm i|b|}{4\beta \hslash^{2}/3}}
\end{equation}
and
\begin{equation}
k'_{\pm}=\sqrt{\frac{1-(|a|\pm i|b|)}{4\beta \hslash^{2}/3}}=\sqrt{\frac{(1-|a|)\mp i|b|}{4\beta \hslash^{2}/3}}.
\end{equation}
We remark again that
\begin{equation}
\sqrt{1\pm (|a|\pm i|b|)}
\end{equation}
is complex number and can be written as
\begin{equation}
z=x+iy.
\end{equation}
Following the same development that led to Eqs. (\ref{+}) and (\ref{-}), we obtain similar results, but with
\begin{equation}
\left\lbrace
\begin{array}{l}
1 \mapsto 1\pm |a|; \\
16m\beta \lambda/3 \mapsto |b|
\end{array}
\right. .
\end{equation}
Thus, we have
\begin{equation}
k_{\pm}=\frac{1}{\sqrt{4\beta \hslash^{2}/3}}\left( 
\sqrt{\frac{\sqrt{(1+|a|)^{2}+|b|^{2}}+(1+|a|)}{2}} \pm i \sqrt{\frac{\sqrt{(1+|a|)^{2}+|b|^{2}}-(1+|a|)}{2}}
\right)
\end{equation}
and
\begin{equation}
k'_{\pm}=\frac{1}{\sqrt{4\beta \hslash^{2}/3}}\left( 
\sqrt{\frac{\sqrt{(1-|a|)^{2}+|b|^{2}}+(1-|a|)}{2}} \mp i \sqrt{\frac{\sqrt{(1-|a|)^{2}+|b|^{2}}-(1-|a|)}{2}}
\right).
\end{equation}
Defining:
$$
a_{\pm}=\frac{\sqrt{(1\pm |a|)^{2}+|b|^{2}}}{4}=\frac{\sqrt{1+|a|^{2}+|b|^{2}\pm 2|a|}}{4},
$$
\begin{equation}
a_{\pm} = \frac{\sqrt{1+\sqrt{1+\left(\frac{16\beta m \lambda}{3}\right)^{2}}\pm \frac{2\sqrt{1+\sqrt{1+\left(\frac{16\beta m \lambda}{3}\right)^{2}}}}{\sqrt{2}}}}{4}
\end{equation}
and
\begin{equation}
b_{\pm}=\frac{1\pm |a|}{4}=\frac{1\pm \frac{\sqrt{1+\sqrt{1+\left(\frac{16\beta m \lambda}{3}\right)^{2}}}}{\sqrt{2}}}{4},
\end{equation}
we find
\begin{equation}
k_{\pm}=\frac{\sqrt{a_{+}+b_{+}}\pm i\sqrt{a_{+}-b_{+}}}{\hslash \sqrt{2\beta /3}} , \qquad k'_{\pm}=\frac{\sqrt{a_{-}+b_{-}}\mp i\sqrt{a_{-}-b_{-}}}{\hslash \sqrt{2\beta /3}}
\end{equation}
and the general solution
\begin{equation}
\label{sol}
\phi_{\pm}=A_{\pm}e^{k_{\pm}x}+B_{\pm}e^{-k_{\pm}x}+C_{\pm}e^{k'_{\pm}x}+D_{\pm}e^{-k'_{\pm}x}.
\end{equation}
Finally, we conclude that\\
\\
I - If $x \in (-\infty,+\infty)$, than $(n_{+},n_{-})=(0,0)$ and, therefore, the Hamiltonian operator is self-adjoint and have no other self-adjoint extension.\\
\\
II - If $x \in [0,+\infty)$, than $(n_{+},n_{-})=(2,2)$ and, therefore, the Hamiltonian operator have infinitely many self-adjoint extensions parametrized by a matrix $U(2)$.\\
\\
III - If $x \in [-a/2,+a/2]$, than $(n_{+},n_{-})=(4,4)$ and, therefore, the Hamiltonian operator have infinitely many self-adjoint extensions parametrized by a matrix $U(4)$.\\

\section{Appendix: Obtaining the $U(4)$ family of self-adjoint boundary conditions}
\label{App. 2}

\h This appendix shows the calculations we performed to obtain the $U(4)$ parametrized boundary conditions (\ref{self-adj. cond.}), that specifies the self-adjoint extensions of the Hamiltonian operator (\ref{corrected ham}). \\
\\
To find the eigenvalues and eigenvectors of the matrix $A$,  Eq. (\ref{A def.}), first we note that
\begin{equation}
A.a= \lambda a \Rightarrow \left( \begin{array}{lr} G & O \\ O & -G \end{array} \right)\left(\begin{array}{lr}a_{+}\\a_{-}\end{array}\right)= \lambda \left(\begin{array}{lr}a_{+}\\a_{-}\end{array}\right) ,
\end{equation}
which implies that 
\begin{equation}
\label{G eq.}
\left\lbrace \begin{array}{ll} G.a_{+}=\lambda a_{+}; \\  G.a_{-}=-\lambda a_{-} .\end{array} \right.
\end{equation}
The top equation of (\ref{G eq.}), together with the definition of $G$, Eq. (\ref{G def.}), gives
\begin{equation}
\label{aut-eq}
\left\lbrace
\begin{array}{rrrr}
-i\,\frac{3a^{2}}{2\beta \hslash^{2}}a_{2} + ia_{4}&=&\lambda a_{1}; \\
i\,\frac{3a^{2}}{2\beta \hslash^{2}}a_{1} - ia_{3}&=&\lambda a_{2}; \\
ia_{2}&=&\lambda a_{3};\\
-ia_{1}&=&\lambda a_{4}.
\end{array}
\right. ,
\end{equation}
where we use
\begin{equation}
a_{+} = \left(\begin{array}{c} a_{1}\\ a_{2} \\ a_{3}
\\ a_{4}\end{array} \right).
\end{equation}
The last two equations of (\ref{aut-eq}) allows us to rewrite the first two in terms only of $a_{1}$ e $a_{2}$:
\begin{equation}
\label{ab1}
-i\frac{3a^{2}}{2\beta \hslash^{2}} \lambda a_{2} + a_{1} = \lambda^{2} a_{1} \, \Rightarrow \, a_{2} = i \frac{2\beta \hslash^{2}}{3a^{2}} . \frac{\lambda^{2}-1}{\lambda}a_{1} 
\end{equation}
and
\begin{equation}
\label{ab2}
i\frac{3a^{2}}{2\beta \hslash^{2}} \lambda a_{1} + a_{2} = \lambda^{2} a_{2} \, \Rightarrow \, a_{2} = i \frac{3a^{2}}{2\beta \hslash^{2}} . \frac{\lambda}{\lambda^{2}-1}a_{1} .
\end{equation}
The product of (\ref{ab1}) and (\ref{ab2}) shows us that
\begin{equation}
a_{2}^{2}= - a_{1}^{2},
\end{equation}
and the ratio of them shows us that
\begin{equation}
\label{lambda eq.}
\left(\frac{2\beta \hslash^{2}}{3a^{2}}\right)^{2}.\left(\frac{\lambda^{2}-1}{\lambda}\right)^{2}=1.
\end{equation}
The solutions of Eq. (\ref{lambda eq.}) give us the eigenvalues of the matrix $G$:
\begin{equation}
\lambda_{1}=+\lambda_{+}, \qquad \lambda_{2}=+\lambda_{-}, \qquad \lambda_{3}=-\lambda_{+}, \qquad \lambda_{4}=-\lambda_{-},
\end{equation}
with $\lambda_{\pm}$ defined by
\begin{equation}
\lambda_{\pm}=\frac{\sqrt{1+\left(\frac{4\beta \hslash^{2}}{3a^{2}}\right)^{2}}\pm 1}{4\beta\hslash^{2}/3a^{2}}.
\end{equation}
It is interesting to note the following properties of the eigenvalues:\\
\\
\\
(i) $\lambda_{+},\lambda_{-}>0$ $\Rightarrow$ $\lambda_{1},\lambda_{2}>0$ and $\lambda_{3},\lambda_{4}<0$;\\
\\
\\
(ii) $\lambda_{+}.\lambda_{-}=1$ $\Rightarrow$ $\lambda_{1}.\lambda_{2}=\lambda_{3}.\lambda_{4}=1$;\\
\\
\\ 
(iii) $\frac{2\beta \hslash^{2}}{3a^{2}}.\frac{\lambda_{\pm}^{2}-1}{\lambda_{\pm}}=\pm 1$ $\Rightarrow$ $\left\lbrace \begin{array}{ll}\frac{2\beta \hslash^{2}}{3a^{2}}.\frac{\lambda^{2}-1}{\lambda}=1, \, $for$ \, \lambda_{1} \, $and$ \, \lambda_{4} \\ \\ \frac{2\beta \hslash^{2}}{3a^{2}}.\frac{\lambda^{2}-1}{\lambda}=-1, \,  $for$ \, \lambda_{2} \, $and$ \, \lambda_{3} \end{array} \right.$.\\
\\
\\
Therefore, by Eqs. (\ref{aut-eq}) and (\ref{ab1}) and the property (iii) above, we can find the following related eigenvalues and eigenvectors of $G$:
\begin{eqnarray}
\label{autval-autvet}
\lambda_{1}: a=\frac{\sqrt{\lambda_{+}}}{\sqrt{2(\lambda_{+}+\lambda_{-})}}\left(\begin{array}{c}1\\i\\-1/\lambda_{1}\\-i/\lambda_{1}\end{array}\right)=\frac{1}{\sqrt{2(\lambda_{+}+\lambda_{-})}}\left(\begin{array}{c}\sqrt{\lambda_{+}}\\i\sqrt{\lambda_{+}}\\-\sqrt{\lambda_{-}}\\-i\sqrt{\lambda_{-}}\end{array}\right) \\ 
\lambda_{2}: b=\frac{\sqrt{\lambda_{-}}}{\sqrt{2(\lambda_{+}+\lambda_{-})}}\left(\begin{array}{c}1\\-i\\1/\lambda_{2}\\-i/\lambda_{2}\end{array}\right)=\frac{1}{\sqrt{2(\lambda_{+}+\lambda_{-})}}\left(\begin{array}{c}\sqrt{\lambda_{-}}\\-i\sqrt{\lambda_{-}}\\ \sqrt{\lambda_{+}}\\-i\sqrt{\lambda_{+}}\end{array}\right) \\
\lambda_{3}: c=\frac{\sqrt{\lambda_{+}}}{\sqrt{2(\lambda_{+}+\lambda_{-})}}\left(\begin{array}{c}1\\-i\\1/\lambda_{3}\\-i/\lambda_{3}\end{array}\right)=\frac{1}{\sqrt{2(\lambda_{+}+\lambda_{-})}}\left(\begin{array}{c}\sqrt{\lambda_{+}}\\-i\sqrt{\lambda_{+}}\\-\sqrt{\lambda_{-}}\\i\sqrt{\lambda_{-}}\end{array}\right) \\
\lambda_{4}: d=\frac{\sqrt{\lambda_{-}}}{\sqrt{2(\lambda_{+}+\lambda_{-})}}\left(\begin{array}{c}1\\i\\-1/\lambda_{4}\\-i/\lambda_{4}\end{array}\right)=\frac{1}{\sqrt{2(\lambda_{+}+\lambda_{-})}}\left(\begin{array}{c}\sqrt{\lambda_{-}}\\i\sqrt{\lambda_{-}}\\ \sqrt{\lambda_{+}}\\i\sqrt{\lambda_{+}}\end{array}\right),
\end{eqnarray}
where the constants were chosen to give $a^{\dagger}.a=b^{\dagger}.b=c^{\dagger}.c=d^{\dagger}.d=1$.\\
\\
Similarly, we note that the bottom equation of (\ref{G eq.}) leads to the same equations expressed in (\ref{aut-eq}), but with $\lambda \mapsto -\lambda$. Thus, we find the same eigenvalues, but with their respective eigenvectors permuted when compared to (\ref{autval-autvet}):
\begin{equation}
\lambda_{1}=+\lambda_{+}: c \qquad \lambda_{2}=+\lambda_{-}: d \qquad  \lambda_{3}=-\lambda_{+}: a \qquad \lambda_{4}=-\lambda_{-}: b .
\end{equation}
Now, we can write the eigenvalues and eigenvectors of the matrix $A$, defined in (\ref{A def.}), as
\begin{equation}
\begin{array}{rr}
\lambda_{1}=+\lambda_{+}: \left(\begin{array}{c} a\\0\end{array}\right)$ and$ \left(\begin{array}{c} 0\\c\end{array}\right) & \lambda_{2}=+\lambda_{-}: \left(\begin{array}{c} b\\0\end{array}\right)$ and$ \left(\begin{array}{c} 0\\d\end{array}\right)  \\ \\ \lambda_{3}=-\lambda_{+}: \left(\begin{array}{c} c\\0\end{array}\right)$ and$ \left(\begin{array}{c} 0\\a\end{array}\right) & \lambda_{4}=-\lambda_{-}: \left(\begin{array}{c} d\\0\end{array}\right)$ and$ \left(\begin{array}{c} 0\\b\end{array}\right)
\end{array},
\end{equation}
and the matrix $A$, itself, as:
\begin{eqnarray}
\label{A diag}
A&=&P^{\dagger}\left(\begin{array}{cccccccc}
\lambda_{+} & 0 & 0 & 0 & 0 & 0 & 0 & 0 \\
0 & \lambda_{+} & 0 & 0 & 0 & 0 & 0 & 0 \\
0 & 0 & \lambda_{-} & 0 & 0 & 0 & 0 & 0 \\
0 & 0 & 0 & \lambda_{-} & 0 & 0 & 0 & 0 \\
0 & 0 & 0 & 0 & -\lambda_{+} & 0 & 0 & 0 \\
0 & 0 & 0 & 0 & 0 & -\lambda_{+} & 0 & 0 \\
0 & 0 & 0 & 0 & 0 & 0 & -\lambda_{-} & 0 \\
0 & 0 & 0 & 0 & 0 & 0 & 0 & -\lambda_{-} \\
\end{array} \right) P \nonumber \\
&=&(D.P)^{\dagger}\left(\begin{array}{cccccccc}
1 & 0 & 0 & 0 & 0 & 0 & 0 & 0 \\
0 & 1 & 0 & 0 & 0 & 0 & 0 & 0 \\
0 & 0 & 1 & 0 & 0 & 0 & 0 & 0 \\
0 & 0 & 0 & 1 & 0 & 0 & 0 & 0 \\
0 & 0 & 0 & 0 & -1 & 0 & 0 & 0 \\
0 & 0 & 0 & 0 & 0 & -1 & 0 & 0 \\
0 & 0 & 0 & 0 & 0 & 0 & -1 & 0 \\
0 & 0 & 0 & 0 & 0 & 0 & 0 & -1 \\
\end{array} \right) (D.P),
\end{eqnarray}
where $P$ is a unitary matrix given by
\begin{eqnarray}
P &=& \frac{1}{\sqrt{2(\lambda_{+}+\lambda_{-})}}\left(\begin{array}{cccccccc}

\sqrt{\lambda_{+}} & -i\sqrt{\lambda_{+}} & -\sqrt{\lambda_{-}} & i\sqrt{\lambda_{-}} & 0 & 0 & 0 & 0 \\

0 & 0 & 0 & 0 & \sqrt{\lambda_{+}} & i\sqrt{\lambda_{+}} & -\sqrt{\lambda_{-}} & -i\sqrt{\lambda_{-}} \\

\sqrt{\lambda_{-}} & i\sqrt{\lambda_{-}} & \sqrt{\lambda_{+}} & i\sqrt{\lambda_{+}} & 0 & 0 & 0 & 0 \\

0 & 0 & 0 & 0 & \sqrt{\lambda_{-}} & -i\sqrt{\lambda_{-}} & \sqrt{\lambda_{+}} & -i\sqrt{\lambda_{+}} \\

\sqrt{\lambda_{+}} & i\sqrt{\lambda_{+}} & -\sqrt{\lambda_{-}} & -i\sqrt{\lambda_{-}} & 0 & 0 & 0 & 0 \\

0 & 0 & 0 & 0 & \sqrt{\lambda_{+}} & -i\sqrt{\lambda_{+}} & -\sqrt{\lambda_{-}} & i\sqrt{\lambda_{-}} \\

\sqrt{\lambda_{-}} & -i\sqrt{\lambda_{-}} & \sqrt{\lambda_{+}} & -i\sqrt{\lambda_{+}} & 0 & 0 & 0 & 0 \\

0 & 0 & 0 & 0 & \sqrt{\lambda_{-}} & i\sqrt{\lambda_{-}} & \sqrt{\lambda_{+}} & i\sqrt{\lambda_{+}}

\end{array} \right) \nonumber \\
&=& \left(\begin{array}{cc}
a^{\dagger} & 0\\
0 & c^{\dagger}\\
b^{\dagger} & 0\\
0 & d^{\dagger}\\
c^{\dagger} & 0\\
0 & a^{\dagger}\\
d^{\dagger} & 0\\
0 & b^{\dagger}
\end{array} \right)
\end{eqnarray} 
and $D$ is a positive diagonal matrix given by
\begin{equation}
D = \left(\begin{array}{cccccccc}
\sqrt{\lambda_{+}} & 0 & 0 & 0 & 0 & 0 & 0 & 0 \\
0 & \sqrt{\lambda_{+}} & 0 & 0 & 0 & 0 & 0 & 0 \\
0 & 0 & \sqrt{\lambda_{-}} & 0 & 0 & 0 & 0 & 0 \\
0 & 0 & 0 & \sqrt{\lambda_{-}} & 0 & 0 & 0 & 0 \\
0 & 0 & 0 & 0 & \sqrt{\lambda_{+}} & 0 & 0 & 0 \\
0 & 0 & 0 & 0 & 0 & \sqrt{\lambda_{+}} & 0 & 0 \\
0 & 0 & 0 & 0 & 0 & 0 & \sqrt{\lambda_{-}} & 0 \\
0 & 0 & 0 & 0 & 0 & 0 & 0 & \sqrt{\lambda_{-}} \\
\end{array} \right).
\end{equation}
With the above definitions, we have:
\begin{equation}
D.P=\frac{1}{\sqrt{2(\lambda_{+}+\lambda_{-})}}\left(\begin{array}{cccccccc}

\lambda_{+} & -i\lambda_{+} & -1 & i & 0 & 0 & 0 & 0 \\

0 & 0 & 0 & 0 & \lambda_{+} & i\lambda_{+} & -1 & -i \\

\lambda_{-} & i\lambda_{-} & 1 & i & 0 & 0 & 0 & 0 \\

0 & 0 & 0 & 0 & \lambda_{-} & -i\lambda_{-} & 1 & -i \\

\lambda_{+} & i\lambda_{+} & -1 & -i & 0 & 0 & 0 & 0 \\

0 & 0 & 0 & 0 & \lambda_{+} & -i\lambda_{+} & -1 & i \\

\lambda_{-} & -i\lambda_{-} & 1 & -i & 0 & 0 & 0 & 0 \\

0 & 0 & 0 & 0 & \lambda_{-} & i\lambda_{-} & 1 & i

\end{array} \right).
\end{equation}

Inserting (\ref{A diag}) in (\ref{sesq form}), and setting it equal to zero, we find

\begin{equation}
\label{mat. formII}
u^{\dagger}.A.v=\left(\begin{array}{c}u_{+} \\ u_{-} \end{array}\right)^{\dagger}.(DP)^{\dagger}.\left( \begin{array}{cc} 1 & 0 \\ 0 & -1 \end{array}\right).(DP).\left(\begin{array}{c}v_{+} \\ v_{-} \end{array}\right)=0,
\end{equation}
which implies that the maximal linear subspaces $V$'s for which (\ref{mat. formII}) is satisfied, for all $u,v\in V$, are given by
\begin{equation}
\label{VU}
V_{U}=\left\lbrace v\in \mathcal{H}|\left(\begin{array}{cc} U & -1 \\ 0 & 0\end{array} \right)(DP)v=0 \right\rbrace ,
\end{equation}
where $U$ is a unitary matrix $4 \times 4$.\\
\\
Finally, we see that $(DP).v$ is given by
\begin{equation}
\frac{1}{\sqrt{2(\lambda_{+}+\lambda_{-})}}\left(\begin{array}{c}

\lambda_{+}\phi(+a/2) - i\lambda_{+}a\partial_{x}\phi(+a/2) - a^{2}\partial_{x}^{2}\phi(+a/2) + ia^{3}\partial_{x}^{3}\phi(+a/2)\\

\lambda_{+}\phi(-a/2) + i\lambda_{+}a\partial_{x}\phi(-a/2) - a^{2}\partial_{x}^{2}\phi(-a/2) - ia^{3}\partial_{x}^{3}\phi(-a/2) \\

\lambda_{-}\phi(+a/2) + i\lambda_{-}a\partial_{x}\phi(+a/2) + a^{2}\partial_{x}^{2}\phi(+a/2) + ia^{3}\partial_{x}^{3}\phi(+a/2) \\

\lambda_{-}\phi(-a/2)  -i\lambda_{-}a\partial_{x}\phi(-a/2) + a^{2}\partial_{x}^{2}\phi(-a/2) - ia^{3}\partial_{x}^{3}\phi(-a/2) \\

\lambda_{+}\phi(+a/2) + i\lambda_{+}a\partial_{x}\phi(+a/2) - a^{2}\partial_{x}^{2}\phi(+a/2) - ia^{3}\partial_{x}^{3}\phi(+a/2) \\

\lambda_{+}\phi(-a/2)  - i\lambda_{+}a\partial_{x}\phi(-a/2) - a^{2}\partial_{x}^{2}\phi(-a/2) + ia^{3}\partial_{x}^{3}\phi(-a/2) \\

\lambda_{-}\phi(+a/2) - i\lambda_{-}a\partial_{x}\phi(+a/2) + a^{2}\partial_{x}^{2}\phi(+a/2) - ia^{3}\partial_{x}^{3}\phi(+a/2) \\

\lambda_{-}\phi(-a/2) + i\lambda_{-}a\partial_{x}\phi(-a/2) + a^{2}\partial_{x}^{2}\phi(-a/2) + ia^{3}\partial_{x}^{3}\phi(-a/2)

\end{array} \right),
\end{equation}
and, therefore, the self-adjointness conditions to $\hat{H}$ are given by
\begin{eqnarray}
\left(\begin{array}{c}

\lambda_{+}\phi(+a/2) + i\lambda_{+}a\partial_{x}\phi(+a/2) - a^{2}\partial_{x}^{2}\phi(+a/2) - ia^{3}\partial_{x}^{3}\phi(+a/2) \\

\lambda_{+}\phi(-a/2)  - i\lambda_{+}a\partial_{x}\phi(-a/2) - a^{2}\partial_{x}^{2}\phi(-a/2) + ia^{3}\partial_{x}^{3}\phi(-a/2) \\

\lambda_{-}\phi(+a/2) - i\lambda_{-}a\partial_{x}\phi(+a/2) + a^{2}\partial_{x}^{2}\phi(+a/2) - ia^{3}\partial_{x}^{3}\phi(+a/2) \\

\lambda_{-}\phi(-a/2) + i\lambda_{-}a\partial_{x}\phi(-a/2) + a^{2}\partial_{x}^{2}\phi(-a/2) + ia^{3}\partial_{x}^{3}\phi(-a/2)

\end{array} \right)
= \nonumber \\
=U\left(\begin{array}{c}

\lambda_{+}\phi(+a/2) - i\lambda_{+}a\partial_{x}\phi(+a/2) - a^{2}\partial_{x}^{2}\phi(+a/2) + ia^{3}\partial_{x}^{3}\phi(+a/2)\\

\lambda_{+}\phi(-a/2) + i\lambda_{+}a\partial_{x}\phi(-a/2) - a^{2}\partial_{x}^{2}\phi(-a/2) - ia^{3}\partial_{x}^{3}\phi(-a/2) \\

\lambda_{-}\phi(+a/2) + i\lambda_{-}a\partial_{x}\phi(+a/2) + a^{2}\partial_{x}^{2}\phi(+a/2) + ia^{3}\partial_{x}^{3}\phi(+a/2) \\

\lambda_{-}\phi(-a/2)  -i\lambda_{-}a\partial_{x}\phi(-a/2) + a^{2}\partial_{x}^{2}\phi(-a/2) - ia^{3}\partial_{x}^{3}\phi(-a/2) \\

\end{array} \right),
\end{eqnarray}
where $U \in U(4)$ is a unitary matrix that specifies the self-adjoint extension. 

\section{Appendix: Solving the infinite square-well}
\label{App. 3}

\h This appendix shows how we went from the boundary conditions (\ref{cond. cont}) to the eigenfunctions (\ref{Asol}) and (\ref{Bsol}) and the energy spectrum conditions (\ref{even spectrum}) and (\ref{odd spectrum}). \\ 
\\
The general solution of the eigenvalue equation
\begin{equation}
\label{H eq.}
\hat{H}\phi(x)= E\phi(x)
\end{equation}
is
\begin{equation}
\label{gen. sol.}
\phi(x) = Ae^{ikx}+Be^{-ikx}+Ce^{k'x}+De^{-k'x}.
\end{equation}
Now, we must impose the boundary conditions (\ref{cond. cont})  in order to get the eigenfunctions (\ref{Asol}) and (\ref{Bsol}) and the energy spectrum conditions (\ref{even spectrum}) and (\ref{odd spectrum}). To this end, we will separate our analysis in the following cases:\\
\\
I - Positive energies $(E>0)$:\\
\\
In this case we have:
\begin{equation}
k=\frac{\sqrt{\sqrt{1+\frac{16}{3}\beta mE}-1}}{2\hslash \sqrt{\beta /3}}, \qquad k'=\frac{\sqrt{\sqrt{1+\frac{16}{3}\beta mE}+1}}{2\hslash \sqrt{\beta /3}}.
\end{equation}
Thus, the conditions (\ref{cond. cont}) take us to the following eigenfunctions of opposite parity\footnote{The constants $A_{k}$ and $B_{k}$ were defined by
\begin{eqnarray}
A_{k}&=&\cos(ka/2)\cosh(k'a/2)\left[ \cosh^{2}(k'a/2) \left(\frac{a}{2}+\frac{(k'^{2}-3k^{2})\sin(ka/2)\cos(ka/2)}{k(k^{2}+k'^{2})}\right) \right. + \nonumber\\
&& + \left. \cos^{2}(ka/2) \left(\frac{a}{2}+\frac{(k^{2}-3k'^{2})\sinh(k'a/2)\cosh(k'a/2)}{k'(k'^{2}+k^{2})}\right) \right]^{-1/2}
\end{eqnarray}
and
\begin{eqnarray}
B_{k}&=&\sin(ka/2)\sinh(k'a/2)\left[ \sinh^{2}(k'a/2) \left(\frac{a}{2}-\frac{(k'^{2}-3k^{2})\sin(ka/2)\cos(ka/2)}{k(k^{2}+k'^{2})}\right) \right. + \nonumber\\
&& - \left. \sin^{2}(ka/2) \left(\frac{a}{2}-\frac{(k^{2}-3k'^{2})\sinh(k'a/2)\cosh(k'a/2)}{k'(k'^{2}+k^{2})}\right) \right]^{-1/2},
\end{eqnarray}
in order to normalize the wave functions.}:
\begin{equation}
\psi_{A_{k}} = A_{k}\left[\frac{\cos(kx)}{\cos(ka/2)}-\frac{\cosh(k'x)}{\cosh(k'a/2)}\right]
\end{equation}
and
\begin{equation}
\psi_{B_{k}} = B_{k}\left[\frac{\sin(kx)}{\sin(ka/2)}-\frac{\sinh(k'x)}{\sinh(k'a/2)}\right],
\end{equation}
as well as the following conditions that define the energy spectrum:
\begin{equation}
k\tan(ka/2)+k'\tanh(k'a/2)=0,
\end{equation}
to the even eigenfunctions, and
\begin{equation}
k\cot(ka/2)-k'\coth(k'a/2)=0
\end{equation}
to the odd ones.\\
\\
II - Null energy $(E=0)$:\\
\\
In this case we have:
\begin{equation}
\label{k', e=0}
k=0, \qquad k'=\frac{\sqrt{2}}{2\hslash \sqrt{\beta /3}}.
\end{equation}
The conditions (\ref{cond. cont}) take us to the following eigenfunctions\footnote{The normalization constants, as well as the eigenfunctions, can be found by taking the limit $k\rightarrow 0$ in the definitions of $A_{k}$ and $B_{k}$, and the results are given, respectively, by
\begin{equation}
A_{0}=\frac{\cosh(k'a/2)}{\sqrt{\frac{a}{2}+a\cosh^{2}(k'a/2)-\frac{3\sinh(k'a/2)\cosh(k'a/2)}{k'}}}
\end{equation}
and
\begin{equation}
B_{0}=\frac{\sinh(k'a/2)}{\sqrt{-\frac{a}{2}+\left(\frac{a}{3}+\frac{8}{ak'^{2}}\right)\sinh^{2}(k'a/2)-\frac{3\sinh(k'a/2)\cosh(k'a/2)}{k'}}}.
\end{equation}
}:
\begin{equation}
\label{Asol2}
\psi_{A_{0}}=A_{0}\left[1-\frac{\cosh(k'x)}{\cosh(k'a/2)}\right]
\end{equation}
and
\begin{equation}
\label{Bsol2}
\psi_{B_{0}}=B_{0}\left[\frac{x}{a/2}-\frac{\sinh(k'x)}{\sinh(k'a/2)}\right],
\end{equation}
as well as the following conditions:
\begin{equation}
\tanh(k'a/2)=0,
\end{equation}
to the even eigenfunctions, and
\begin{equation}
\tanh(k'a/2)=k'a/2
\end{equation}
to the odd ones. Since, for $a\neq 0$, those conditions could only be satisfied if $k'=0$, which contradicts (\ref{k', e=0}), we conclude that this self-adjoint extension does not admit null energy solutions.\\
\\
III - Negative energies I $(-3/16m\beta <E<0)$:\\
\\
In this case, we have:
\begin{equation}
\label{k e<0}
k=\frac{1}{2\hslash \sqrt{\beta /3}}\sqrt{1-\sqrt{1-\frac{16}{3}m\beta |E|}}
\end{equation}
and
\begin{equation}
\label{k' e<0}
k'=\frac{1}{2\hslash \sqrt{\beta /3}}\sqrt{1+\sqrt{1-\frac{16}{3}m\beta |E|}}.
\end{equation}
The boundary conditions (\ref{cond. cont}) take us to the following eigenfunctions\footnote{The new constants $A_{k}$ and $B_{k}$, as well as the eigenfunctions itself, can be obtained by the substitution $k\mapsto ik$, and can be written explicitly as
\begin{eqnarray}
A_{k}&=&\cosh(ka/2)\cosh(k'a/2)\left[ \cosh^{2}(k'a/2) \left(\frac{a}{2}+\frac{(k'^{2}+3k^{2})\sinh(ka/2)\cosh(ka/2)}{k(k'^{2}-k^{2})}\right) \right. + \nonumber\\
&& + \left. \cosh^{2}(ka/2) \left(\frac{a}{2}+\frac{(k^{2}+3k'^{2})\sinh(k'a/2)\cosh(k'a/2)}{k'(k^{2}-k^{'2})}\right) \right]^{-1/2}
\end{eqnarray}
and
\begin{eqnarray}
B_{k}&=&\sinh(ka/2)\sinh(k'a/2)\left[- \sinh^{2}(k'a/2) \left(\frac{a}{2}-\frac{(k'^{2}+3k^{2})\sinh(ka/2)\cosh(ka/2)}{k(k'^{2}-k^{2})}\right) \right. + \nonumber\\
&& - \left. \sinh^{2}(ka/2) \left(\frac{a}{2}-\frac{(k^{2}+3k'^{2})\sinh(k'a/2)\cosh(k'a/2)}{k'(k^{2}-k'^{2})}\right) \right]^{-1/2}.
\end{eqnarray}
}:
\begin{equation}
\label{Asol3}
\psi_{A_{k}}=A_{k}\left[\frac{\cosh(kx)}{\cosh(ka/2)}-\frac{\cosh(k'x)}{\cosh(k'a/2)}\right]
\end{equation}
and
\begin{equation}
\label{Bsol3}
\psi_{B_{k}}=B_{k}\left[\frac{\sinh(kx)}{\sinh(ka/2)}-\frac{\sinh(k'x)}{\sinh(k'a/2)}\right],
\end{equation}
as well as to the following conditions:
\begin{equation}
k\tanh(ka/2)=k'\tanh(k'a/2),
\end{equation}
to the even eigenfunctions, and
\begin{equation}
k\coth(ka/2)=k'\coth(k'a/2)
\end{equation}
to the odd ones. Since $a \neq 0$, those conditions could only be satisfied if $k=k'$, what is not allowed by Eqs. (\ref{k e<0}) and (\ref{k' e<0}) in the present interval of energy. Thus, we conclude that this self-adjoint extension of $\hat{H}$ does not admit negative energy solutions greater than $-3/16m\beta$\\
\\
IV - Negative energy II $(E=-3/16m\beta)$:\\
\\
In this case, we have:
\begin{equation}
\label{k' e=-}
k=k'=\frac{1}{2\hslash \sqrt{\beta /3}}.
\end{equation}\\
The boundary conditions (\ref{cond. cont}) take us to the following eigenfunctions\footnote{Both the eigenfunctions and the normalization constants can be obtained by taking the limit $k\rightarrow ik'$ in (\ref{Asol}) and (\ref{Bsol}) or taking the limit $k\rightarrow k'$ in (\ref{Asol3}) and (\ref{Bsol3}). In both cases the limit has to be taken with the constants included, which in turn are given now by
\begin{equation}
\label{Asol4}
A=\sinh(ka/2)\cosh(ka/2)\left[-\frac{a}{2}+\cosh^{2}(ka/2)\left(\frac{2a}{3}-\frac{1}{2(a/2)k^{2}}+\frac{\sinh(ka/2)\cosh(ka/2)}{2(a/2)^{2}k^{3}}\right) \right]^{-1/2}
\end{equation}
and
\begin{equation}
\label{Bsol4}
B=\sinh(ka/2)\cosh(ka/2)\left[-\frac{a}{2}+\sinh^{2}(ka/2)\left(\frac{2(a/2)}{3}-\frac{1}{2(a/2)k^{2}}+\frac{\sinh(ka/2)\cosh(ka/2)}{2(a/2)^{2}k^{3}}\right) \right]^{-1/2}
\end{equation}
}:
\begin{equation}
\psi_{A}=A\left[\frac{\cosh(k'x)}{\cosh(k'a/2)}-\frac{x}{a/2}\frac{\sinh(k'x)}{\sinh(k'a/2)}\right]
\end{equation}
and
\begin{equation}
\psi_{B}=B\left[\frac{\sinh(k'x)}{\sinh(k'a/2)}-\frac{x}{a/2}\frac{\cosh(k'x)}{\cosh(k'a/2)}\right],
\end{equation}
as well as to the following conditions:
\begin{equation}
\frac{k'a}{2}\left[\tanh(k'a/2)-\coth(k'a/2)\right]=1,
\end{equation}
to the even eigenfunctions, and
\begin{equation}
\frac{k'a}{2}\left[\tanh(k'a/2)-\coth(k'a/2)\right]=-1
\end{equation}
to the odd ones. After some algebra, we can rewrite the above conditions in the form:
\begin{equation}
\frac{\sinh(k'a)}{k'a}=\mp 1,
\end{equation}
whose only solution, for $a \neq 0$, would be $k'=0$. Since this solution contradicts (\ref{k' e=-}), we conclude that this self-adjoint extension of $\hat{H}$ does not admit solutions with energy $-3/16m\beta$ either.
\\
\\
V - Negative Energy III $(E<-3/16m\beta)$:\\
\\
In this case we have:
\begin{equation}
\label{k e<-}
k=\frac{1}{2\hslash \sqrt{\beta /3}}\sqrt{\frac{\sqrt{\frac{16}{3}m\beta |E|}-1}{2}}
\end{equation}
and
\begin{equation}
\label{k' e<-}
k'=\frac{1}{2\hslash \sqrt{\beta /3}}\sqrt{\frac{\sqrt{\frac{16}{3}m\beta |E|}+1}{2}}.
\end{equation}
The boundary conditions (\ref{cond. cont}) take us to the following eigenfunctions:
\begin{equation}
\label{Asol5}
\psi_{A_{k}}=A_{k}\left[\frac{\cos(kx).\cosh(k'x)}{\cos(ka/2).\cosh(k'a/2)}-\frac{\sin(kx).\sinh(k'x)}{\sin(ka/2).\sinh(k'a/2)}\right]
\end{equation}
and
\begin{equation}
\label{Bsol5}
\psi_{B_{k}}=B_{k}\left[\frac{\cos(kx).\sinh(k'x)}{\cos(ka/2).\sinh(k'a/2)}-\frac{\sin(kx).\cosh(k'x)}{\sin(ka/2).\cosh(k'a/2)}\right],
\end{equation}
as well as the following conditions:
\begin{equation}
k\left[\tan(ka/2)+\cot(ka/2)\right]= k'\left[\tanh(k'a/2)-\coth(k'a/2)\right],
\end{equation}
to the even eigenfunctions, and
\begin{equation}
k\left[\tan(ka/2)+\cot(ka/2)\right]=- k'\left[\tanh(k'a/2)-\coth(k'a/2)\right]
\end{equation}
to the odd ones. Once again, the above conditions can be written in a simplified form as:
\begin{equation}
\frac{\sin(ka)}{ka}=\mp \frac{\sinh(k'a)}{k'a},
\end{equation}
whose only solution, for $a \neq 0$, is given by $k=k'=0$. Since this solution contradicts Eqs. (\ref{k e<-}) and (\ref{k' e<-}), we conclude that this self-adjoint extension does not admit energy less than $-3/16m\beta$ either.

\section{Appendix: Symmetry properties of some solutions}
\label{App. 4}

\h This appendix shows the interesting properties of the solutions of Section \ref{SOMFDCOTHSAE} under parity, time inversion and translation.

\subsection{Periodic conditions}

If $\hat{P}$ and $\hat{T}$ are respectively the operators of parity\footnote{In this case we define parity in relation to the point $\delta=a.\theta /(2n\pi)$, not to the origin, as usually. It can be done defining the operator $\hat{P}(\delta)$ that, if $f(x)$ are in the domain of $\hat{P}(\delta)$, $\hat{P}(\delta)f(x)=f(2\delta-x)$ -- note that, indeed, $\hat{P}(\delta)f(x-\delta)=f(\delta-x)$. If $\delta=0$, we turn back to the conventional definition of parity. If parity were defined in relation to the origin, the Eq. (\ref{P T action II}) would be correct except for a phase factor and all the following discussion would still be valid.} and time inversion\footnote{Note that the phases in (\ref{ring sol.}) and (\ref{ring sol. II}) were chosen to give the results (\ref{P T action}) and (\ref{P T action II}) to the operator $\hat{T}$. For any other choice, those results would be the same, except for a phase factor, and all the following discussion would still be valid.}, we have
\begin{equation}
\label{P T action}
\hat{P}\phi_{0}(x)= \hat{T}\phi_{0}(x)=\phi_{0}(x)
\end{equation}
and
\begin{equation}
\label{P T action II}
\hat{P}\phi_{n}^{\pm}(x)= \hat{T}\phi_{n}^{\pm}(x)=\phi_{n}^{\mp}(x).
\end{equation}
Since
\begin{equation}
\hat{H}\phi_{0}(x)=E_{0}\phi_{0}=0
\end{equation}
and
\begin{equation}
\hat{H}\phi_{n}^{\pm}(x)=E_{n}\phi_{n}^{\pm}(x),
\end{equation}
we conclude that the operators $\hat{H}$, $\hat{P}$ e $\hat{T}$ commute with each other. Thus, since the ground state is invariant under parity and time inversion, i.e., $\phi_{0}(x)$ is an eigenfunction of $\hat{P}$ and $\hat{T}$ (\ref{P T action}), we say that those symmetries are not spontaneously broken \cite{Capri:1977}. Besides that, the solutions (\ref{ring sol.}) are not symmetric by parity and time inversion, except for the ground state, since they are not invariant under the action of the operators $\hat{T}$ and $\hat{P}$ -- according to the Eq. (\ref{P T action II}). Although the solutions (\ref{ring sol.}) are not symmetric under parity and time inversion, they have translational symmetry, in the sense that the probability of find the particle in a interval $0 < \Delta x < a$ around some point $x$, given by
\begin{equation}
\int_{x-\Delta x/2}^{x+\Delta x/2}|\phi(x)|^{2}dx= \frac{\Delta x}{a},
\end{equation}
is not dependent on $x$ (although it does depend upon the size of the interval). These proprieties can be properly understood if we note that the degenerate solutions (\ref{ring sol.}) represent waves traveling to one or the opposite direction on the ring, but with equal probability of being found at any place of it. We can rebuilt the symmetries of parity and time inversion introducing the following change of base:
\begin{eqnarray}
\label{ring new base}
\psi^{+}_{n}(x)=\frac{1}{\sqrt{2}}\left[\phi^{+}_{n}(x)+\phi^{-}_{n}(x)\right] 
=\sqrt{\frac{2}{a}}\, \cos \left(\frac{2n\pi x}{a}-\theta\right),\\
\label{ring new base II}
\psi^{-}_{n}(x)=\frac{-i}{\sqrt{2}}\left[\phi^{+}_{n}(x)-\phi^{-}_{n}(x)\right] 
=\sqrt{\frac{2}{a}}\, \sin \left(\frac{2n\pi x}{a}-\theta\right).
\end{eqnarray}
In this new base, we see that\footnote{Eqs. (\ref{ring new base}) and (\ref{ring new base II}) are defined only for $n \in \mathbb{N}^{*}$. To the ground state, we still have (\ref{ring sol. II}) and, therefore, (\ref{P T action}).}
\begin{eqnarray}
\label{ring new relation}
\hat{P}\psi^{\pm}_{n}(x)&=&\pm \psi^{\pm}_{n}(x);\\
\hat{T}\psi^{\pm}_{n}(x)&=&\psi^{\pm}_{n}(x);\\
H\psi^{\pm}_{n}(x)&=&E_{n}\psi^{\pm}_{n}(x)
\end{eqnarray}
and, therefore, not only the operators $\hat{P}$, $\hat{T}$ and $\hat{H}$ commute with each other, but also have the solutions (\ref{ring new base}) and (\ref{ring new base II}) as eigenfunctions, which guarantees the symmetry under parity and time inversion. On the other hand, we see now that the symmetry under translation was broken (except for the ground state, which may represent a particle in rest at some place on the ring), what can be easily seen if we note that the density of probability $|\psi^{\pm}_{n}(x)|^{2}$ is a function of $x$ and, therefore, the probability of finding the particle in a small region $0<\Delta x<a$ around $x$ varies with $x$. Now, we can interpret the new base solutions (\ref{ring new base}) as stationary waves on the ring, which are naturally symmetric under parity and time inversion, but not under translation, since the probability of finding the particle is null at the nodes and maximum at the antinodes. Finally, we remember that the constant $\theta$ was introduced in (\ref{ring sol.}) to single out the non-special character of the origin on a ring. It is also worth to note that, if we take $\theta=0$, we will not return to the solutions (\ref{ord. sol.}). Although $\psi^{-}$ coincides with the odd solutions of that case, the same is not true for $\psi^{+}$ and for the even solutions.

\subsection{Anti-periodic conditions}

Applying $\hat{P}$ and $\hat{T}$ to the eigenstates (\ref{anti-ring sol.}), wee see that
\begin{equation}
\label{P T anti action}
\hat{P}\phi_{n}^{\pm}(x)= \hat{T}\phi_{n}^{\pm}(x)=\phi_{n\pm 1}^{\mp}(x).
\end{equation}
Since
\begin{equation}
\hat{H}\phi_{n}^{\pm}(x)=E_{n}^{\pm}\phi_{n}^{\pm}(x)=E_{n\pm 1}^{\mp}\phi_{n}^{\pm}(x),
\end{equation}
we conclude that the operators $\hat{H}$, $\hat{P}$ and $\hat{T}$ commute with each others. Besides the commutations, since none of the $\hat{H}$ eigenstates (\ref{anti-ring sol.}) are invariant under $\hat{P}$ or $\hat{T}$ -- that is, none of them is a eigenstate of $\hat{P}$ or $\hat{T}$ (not even the ground state) -- we say that those symmetries are spontaneously broken. As in the periodic case, even though  (\ref{anti-ring sol.}) are not symmetric under parity and time inversion, they are symmetric under translation in the same sense we mentioned before. Again, we can rebuild the symmetries under parity and time inversion introducing the following change of basis:
\begin{eqnarray}
\label{anti ring new base}
\psi^{+}_{n}(x)=\frac{1}{\sqrt{2}}\left[\phi^{+}_{n}(x)+\phi^{-}_{n+1}(x)\right] 
=\sqrt{\frac{2}{a}}\, \cos \left[\frac{(2n+1)\pi x}{a}-\theta\right],\\
\label{anti ring new base II}
\psi^{-}_{n}(x)=\frac{-i}{\sqrt{2}}\left[\phi^{+}_{n}(x)-\phi^{-}_{n+1}(x)\right] 
=\sqrt{\frac{2}{a}}\, \sin \left[\frac{(2n+1)\pi x}{a}-\theta\right].
\end{eqnarray}
On this new base, we see that:
\begin{eqnarray}
\label{anti ring new relation}
\hat{P}\psi^{\pm}_{n}(x)&=&\pm \psi^{\pm}_{n}(x);\\
\hat{T}\psi^{\pm}_{n}(x)&=&\psi^{\pm}_{n}(x);\\
\hat{H}\psi^{\pm}_{n}(x)&=&E_{n}\psi^{\pm}_{n}(x)
\end{eqnarray}
and, therefore, not only the operators $\hat{P}$, $\hat{T}$ and $\hat{H}$ commute with each others, but have the solutions (\ref{anti ring new base}) and (\ref{anti ring new base II}) as eigenfunctions, ensuring the symmetry under parity and time inversion. On the other hand, we had to give up the symmetry of translation, which was broken again (even to the ground states). Finally, we note that, for $\theta=0$, we do not return to the solutions (\ref{ord. sol.}) either, but, in this case, we see that are the solutions $\psi^{-}$ which doesn't coincides with the odd solutions of that case, since the solutions $\psi^{+}$ do coincide with the even ones.

\subsection{Periodic conditions up to a phase factor}

Unlike the previous cases, it is not hard to see that the operators $\hat{P}$ and $\hat{T}$ do not commute with $\hat{H}$. Indeed, we see that the action of $\hat{P}$ and $\hat{T}$ on (\ref{phase ring sol.}) leads us to wave functions that do not satisfy (\ref{phase ring cond.}) and, thus, are not in the domain of $H$. We also note that no change of basis can restore the symmetries under parity and time inversion. Such results were already expected, because the presence of $e^{i\varphi}$ in the boundary conditions breaks its parity symmetry -- even though it is still present in the potential and in the Hamiltonian -- and also prevents the complex conjugate of any solution $\phi_{n}(x)$ to satisfy (\ref{phase ring cond.}) -- which is a necessary condition to be invariant under time inversion. In the case of parity, it is enough to note that the boundary conditions have definite parity if, and only if, $\varphi=0$ or $\varphi=\pi$. Similarly, for time inversion, taking the complex conjugate of (\ref{phase ring cond.}) we are led to,
\begin{equation}
\label{complex phase ring cond.}
\left\lbrace \begin{array}{c}
\phi^{*}(+a/2)=e^{-i\varphi}.\phi^{*}(-a/2)\\
\partial_{x}\phi^{*}(+a/2)=e^{-i\varphi}.\partial_{x}\phi^{*}(-a/2)\\
\partial_{x}^{2}\phi^{*}(+a/2)=e^{-i\varphi}.\partial_{x}^{2}\phi^{*}(-a/2)\\
\partial_{x}^{3}\phi^{*}(+a/2)=e^{-i\varphi}.\partial_{x}^{3}\phi^{*}(-a/2)
\end{array} \right. ,
\end{equation}
from which we conclude that, except for $\varphi=0$ or $\varphi=\pi$, $\phi^{*}(x)$ does not satisfy the same boundary conditions that $\phi(x)$ and, hence, cannot be a solution when $\phi(x)$ is.





\end{document}